\RequirePackage{fix-cm}
\documentclass[smallextended]{svjour3}       % onecolumn (second format)
\smartqed  % flush right qed marks, e.g. at end of proof
\usepackage{graphicx}
\usepackage[numbers]{natbib}

\PassOptionsToPackage{hyphens}{url}\usepackage{hyperref}

\begin{document}

\title{Multiscale method for Oseen problem in porous media with non-periodic grain patterns%\thanks{Grants or other notes
%about the article that should go on the front page should be
%placed here. General acknowledgments should be placed at the end of the article.}
}
%\subtitle{Crouzeix-Raviart MsFEM for \\ Oseen problem}

\titlerunning{Non-conforming MsFEM for Oseen problem}        % if too long for running head

\author{Bagus Putra Muljadi}

%\authorrunning{Short form of author list} % if too long for running head

\institute{B.P. Muljadi \at
              Department of Earth Science and Engineering, \\
              Imperial College, London, \\
              SW7 2BP, United Kingdom \\
              Tel.: +44 7475834586\\
              \email{b.muljadi@imperial.ac.uk}           %  \\
%             \emph{Present address:} of F. Author  %  if needed
}

\date{Received: \today / Accepted: }
% The correct dates will be entered by the editor

\maketitle

\begin{abstract}
Accurate prediction of the macroscopic flow parameters needed to describe flow in porous media relies on a good knowledge of flow field distribution at a much smaller scale---in the pore spaces. The extent of the inertial effect in the pore spaces can not be underestimated yet is often ignored in large-scale simulations of fluid flow. We present a multiscale method for solving Oseen's approximation of incompressible flow in the pore spaces amid non-periodic grain patterns. The method is based on the multiscale finite element method (MsFEM \cite{thomhou}) and is built in the vein of Crouzeix-Raviart elements \cite{CRRairo}.  Simulations of inertial flow in highly non-periodic settings are conducted and presented. Convergence studies in terms of numerical errors relative to the reference solution are given to demonstrate the accuracy of our method. The weakly enforced continuity across coarse element edges is shown to maintain accurate solutions in the vicinity of the grains without the need for any oversampling methods. The penalization method is employed to allow a complicated grain pattern to be modeled using a simple Cartesian mesh. This work is a stepping stone towards solving the more complicated Navier-Stokes equations with a non-linear inertial term.
\keywords{Crouzeix-Raviart element \and Oseen approximation \and Multiscale finite element method \and Penalization method}
% \PACS{PACS code1 \and PACS code2 \and more}
% \subclass{MSC code1 \and MSC code2 \and more}
\end{abstract}

\section{Introduction}
Modeling of flow through porous bodies is a topic of high importance in various fields of engineering, chemical, biological or geological applications. One of the most significant challenges persisting in virtually all these areas is the disparity between the spatial scales at which flow and transport can be understood; and the scales at which practical model predictions are needed \cite{scheibe}. This disparity in scales forces a trade-off between building models which suffice for practical application, and models that solve the problem \textit{ab-initio} but which may not be able to cope with large-scale problems adequately. To place this in context, in geological media, X-ray techniques now allow three-dimensional images to be acquired routinely \cite{Blunt2013image}.  The pore spaces of these rocks are typically of order microns across.  However, for practical applications in oil recovery, carbon dioxide storage and contaminant transport, flow over 100s m to km needs to be predicted.  This enormous range of scales precludes the use of a direct method that resolves pore-scale flow while determining reservoir-scale behavior.  Instead, techniques that can approximate the flow over distances much larger than the pore scale are needed. A number of multiscale simulation paradigms have been developed to bridge first-principles and empirical methods, and provide a link between micro, and macroscale models.  An additional problem is that in many applications, such as flow in fractured rock and near-well bore flows, the non-linear, or inertial term in the Navier-Stokes equation are significant.  This means that at the large-scale, the application of the linear Darcy-law for flow is inaccurate. 

Our choice of a particular multiscale method is based on the following. First, we consider direct Navier-Stokes simulation on the pore geometry as the holy-grail of microscale simulation for it is considered to be the most complex, and highly resolved spatially (although Navier-Stokes itself can be seen as an upscaled representation of molecular-scale interactions, with effective parameters such as viscosity and density). Such a microscale model strikes a balance between the appropriate level of complexity with current technological advances. For example, recent developments in both computational algorithms, and increases in computer power, coupled with the availability of pore-scale images, have enabled the routine prediction of permeability, with direct Navier-Stokes calculations on samples containing up to a billion voxels \cite{Blunt2013image,Mostaghimi2012}. Second, we are interested in a method capable of resolving the microscale model directly over the domain of interest without losing any degrees of freedom---which rules out other multiscale methods which borrow their philosophy from homogenisation theory (e.g  formal upscaling with closure approximation). Several multiresolution solvers are designed for this purpose, i.e to provide computationally efficient ways of obtaining a complete solution on the fine grid, for example: multigrid solvers and preconditioners \cite{wesseling1992introduction}, multiscale finite element methods (MsFEM) \cite{thomhou,Aarnes2005257,Jenny200347}, and multiscale mimetic methods \cite{Lipnikov}. We choose to develop an adaptation of MsFEM dedicated for solving flow in a pore domain left void by non-periodic grain patterns, which is a representation of all natural pore structures.

The challenge in applying MsFEM in a non-periodic setting is to avoid an intersection between a coarse element boundary and a grain. On the other hand, the overall performance of MsFEM rely on the accuracy of the multiscale basis function which is very sensitive the treatment of subgrid boundary condition. The application of oversampling methods \cite{efendievetal,chuetal,henningpeterseim} was intended to circumvent this problem by broadening the domain in which basis functions are sampled. While the methods perform satisfactorily in the context of perforated media \cite{lozinskibubbleetal,Guanglian}, nevertheless it necessitates an \textit{ad hoc} parameterisation and results in a larger computational problem. Another alternative is to adopt a nonconforming finite element method and impose only a \textit{weak} continuity between coarse element boundaries and therefore allowing the coarse element boundaries to adapt to random patterns of grains. In our previous works, the nonconforming Crouzeix-Raviart element has been adopted successfully for solving advection-diffusion and Stokes equations \cite{clbetal,muljadicrmsfem,MuljadiSIAM}.

In the context of flow pass porous bodies, creeping or Stokes flow is often assumed. This ceases to apply, as mentioned above, for example, near propped fractures, or boreholes in reservoirs where inertial forces becomes dominant. Even in the absence of fractures, Muljadi et al \cite{Muljadi2015} studied the non-Darcy flow behaviour in porous media with different pore heterogeneities and found that the cessation of the Darcy relationship in Estaillades limestone already takes place at Re $\approx 0.001$, three orders of magnitude smaller than what suggested in the literature (Re $\approx1$) based on studies of homogeneous media, such as bead packs, which are poor representations of the heterogeneous reservoir rocks of practical interest. The difficulty in solving the full Navier-Stokes equation is the non-linear nature of its inertial term. As a stepping stone towards appying MsFEM on the more complicated Navier-Stokes equation, we present the framework for solving Oseen's approximation of incompressible flow which provides a linearisation of the inertial term. Note that when solving the full Navier-Stokes equation, often discrete Oseen's problems are solved iteratively in each time step.

To avoid having to work with complicated boundary fitted or even unstructured meshes, we employ the penalization method \cite{Bruneau} when modeling non-periodic grain patterns. Here we simply \textit{force} the solution to vanish within the grain boundaries. Consequentially, this approach allows the modeling of a complicated grains pattern on a simple Cartesian mesh.

This paper is organised as follows. The formulation of the problem is given in section \ref{sec:formulation}. The construction of Crouzeix-Raviart elements is presented in section \ref{sec:crmsfem}. In section \ref{sec:penalization}, the application of penalization method on our problem is described. Then the description of the computation of the reference solutions is given in section \ref{sec:reference} followed by some remarks on the treatment of the boundary condition in section \ref{sec:bc}. Numerical tests are presented and the results discussed in section \ref{sec:numerical} followed by some concluding remarks.

%%%%%%%%%%%%%%%%%%%%%%%%%%%%%%%%%%%%%%FIGURES%%%%%%%%%%%%%%%%%%%%%%
\begin{figure}[hbt]
	\centering
	\includegraphics[width = 0.55\linewidth]{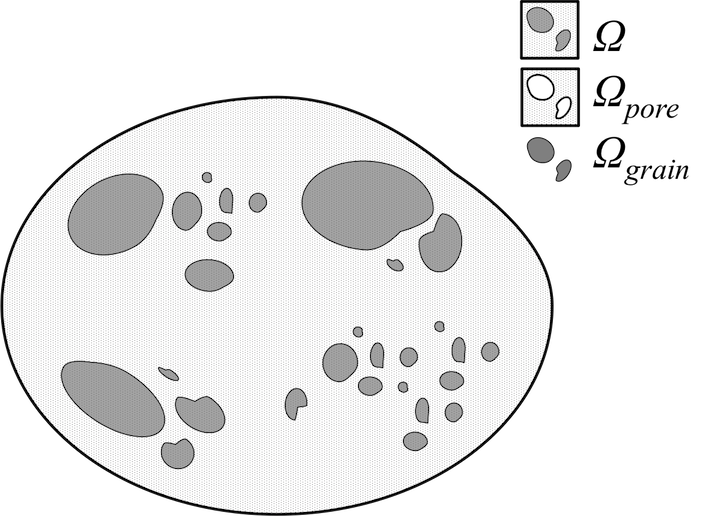} 
	\caption{An illustration of a domain $\Omega$ consisting of a pore domain $\Omega_\textrm{pore}$, and a grain domain $\Omega_\textrm{grain}$.}
	\label{fig:domain}
\end{figure} 
%%%%%%%%%%%%%%%%%%%%%%%%%%%%%%%%%%%%%%%%%%%%%%%%%%%%%%%%%%%%%%%%%%%

\section{Problem Formulation}
\label{sec:formulation} %rewritte and simplify
We define $\Omega$, a two-dimensional domain consisting of a grain domain $\Omega_\textrm{grain}$, and a pore domain $\Omega_\textrm{pore}$ perforated by grains, see figure \ref{fig:domain}. Then, let $\varepsilon$ denote the diameter of the smallest grain. The steady-state Oseen's problem is to find velocity $\vec{u}$ which is the solution to:
\begin{eqnarray} 
\mu\nabla^2\vec{u}+\rho(\vec{U}\cdot\nabla\vec{u})-\nabla p = \vec{f} & \hspace{5mm}\textrm{in}\hspace{5mm} & \Omega_{\textrm{pore}}\label{eq:main}\\
\nabla \cdot \vec{u} = 0 & \hspace{5mm}\textrm{in}\hspace{5mm} & \Omega_{\textrm{pore}} \nonumber
\end{eqnarray}
where $\mu$ is the dynamic viscosity, $\rho$ is the density and $\vec{U}$ is a known velocity field. 

The boundary condition of equation (\ref{eq:main}) is given by:
\begin{eqnarray}
%\vec{u}=0 & \hspace{5mm}\textrm{on}\hspace{5mm} & \partial \Omega_{\textrm{grain}} \cap \partial\Omega_{\textrm{pore}}\\
\vec{u}&=&\vec{g} \textrm{,  on   }  \partial \Omega \cap \partial\Omega_{\textrm{pore}}, \\
\vec{u} &=& \vec{U} \textrm{, when } \{x,y\}\rightarrow\infty\nonumber,
\end{eqnarray}
where $\vec{f}$ is a source function, and $\vec{g}$ is a function fixed at the boundary $\partial\Omega$. In this paper, we consider only a no-slip condition on the grain boundaries: $\vec{u}\textrm{ at }{\partial \Omega_{\textrm{grain}}} = 0$.

\begin{figure}[hbt]
	\centering
	\includegraphics[width = 0.65\linewidth]{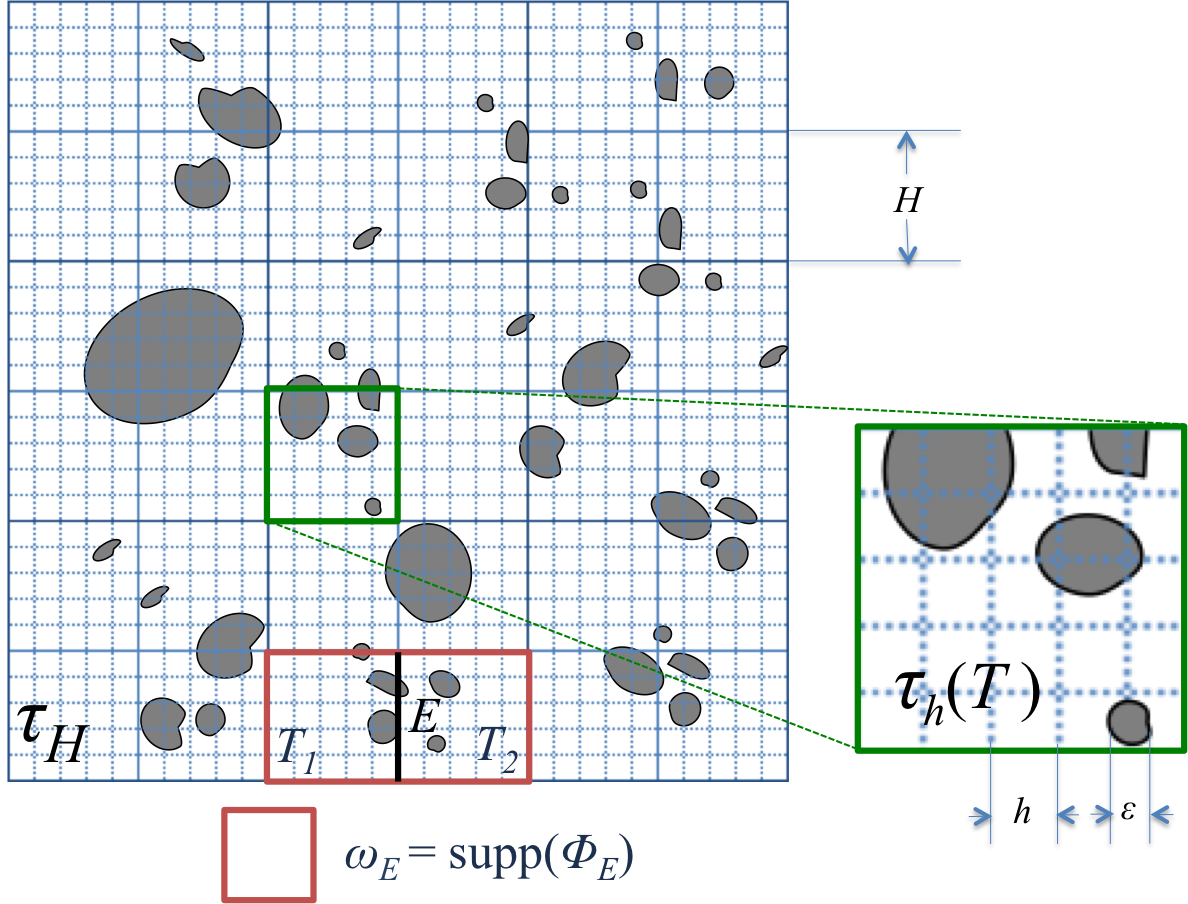} 
	\caption{An illustration of the discretised domain $\tau_H$, and $\omega _{E}$ which is the support space for $\vec{\Phi}_{E}$.}
	\label{fig:compDom}
\end{figure}
\section{Application of Crouzeix-Raviart MsFE}
\label{sec:crmsfem}
Here we explain the application of our method starting from the definition of the coarse and fine meshes. We then introduce the functional spaces for our multiscale basis functions and describe the construction of these functions within each coarse elements. 
\subsection{Discretisation}
We discretise $\Omega$ into a two-dimensional homogeneous Cartesian coarse mesh $\mathcal{T}_H$ (see figure \ref{fig:compDom}). $\mathcal{T}_H$ consists of coarse elements $T_k, k=1,2,\dots,{N_H}$, where $N_H$ is the total number of coarse elements, each with width $H$. We define $\mathcal{E}_H$ the set of all coarse edges $E_j,j=1,2,\dots,{N_E}$ in $\mathcal{T}_H$ including the edges on the domain boundary $\partial\Omega$. For each element $T$ we construct a fine mesh $\mathcal{T}_h(T)$, consisting of fine elements each with width $h$. Note that the combination of $\mathcal{T}_h(T)$ for all $T\in\mathcal{T}_H$ constructs a global fine mesh $\mathcal{T}_h$ which overlaps with $\mathcal{T}_H$. Conversely, one can generate $\mathcal{T}_H$ from $\mathcal{T}_h$ since the difference between the two meshes is only in the indexing. 

\subsection{Crouzeix--Raviart functional spaces}
The functional spaces for velocity $V_H$, and for pressure $M_H$ are given below: 
\begin{eqnarray}
\label{eq:funcSpace}
M_H &=& \{{q}\in L^2 \, \mbox{such that} \, q = \mbox{constant, }\forall T \in \mathcal{T}_H \},\\
V_H &=& \{\vec{v}:\Omega \rightarrow R^2 \, : \, \forall T \in \mathcal{T}_H \,\mbox{such that}\nonumber\\
&& \mu \nabla^2 \vec{v} + \rho(\vec{U}\cdot\nabla\vec{v}) -\nabla s = 0 \textrm{, in }\Omega_\textrm{pore}\cap T  \nonumber\\
&& \nabla\cdot\,\vec{v} = \textrm{ constant in }\Omega_\textrm{pore}\cap T\nonumber\\
&&\nabla \vec{v}\,n-sn=\textrm{ constant on }E\cap \Omega_\textrm{pore},
\  \forall E\in\mathcal{E}(T)\}.\nonumber
%&& {n\cdot\mu\nabla {\vec{v}}_{K}}_{|_e} = c_{K,e}\footnotemark, \hspace{2mm} K = 1,2 \nonumber\\
%&& \mbox{for all edges} \, E \in \varepsilon_H\}. \nonumber
\end{eqnarray}
%\footnotetext{$c_{K,e}$ a constant}
%where $c$ is a constant. %Note that if $\nabla\cdot\vec{v}=0$ in $T$, then $\int_{\partial T}\vec{v}\cdot n = 0$. 
The key here is to maintain the continuity of (only) the average of $\vec{v}$ across an edge $E$: $\int_E [[\vec{v}]]=0$, where $[[{v}]]$ is the \textit{jump} of the value $v$ across $E$. We wish to retain the advantage of our approach which has been successfully applied on Advection-Diffusion, and Stokes problems, namely: the weak imposing of continuity across element boundaries allows adaptive boundary conditions which relaxes the sensitivity of our method to random arrangements of grains, without the need of applying the more cumbersome oversampling methods.

\subsection{Coarse-scale solution}
By discretising $p$ into $p_H$ and $\vec{u}$ into $\vec{u}_H$, we can solve equation (\ref{eq:main}) in $\tau_H$ and rewrite it in a weak form as:
\begin{eqnarray}
a(\vec{u}_H,\vec{v}_H) + c(\vec{u}_H,\vec{v}_H) + b(\vec{v}_H,p_H) = &(\vec{v}_H,\vec{f}) &,\forall \vec{v}_H \in V_H \label{maindiscret} \\
b(\vec{u}_H,q_H) = &0 &,\forall q_H \in M_H \nonumber
\end{eqnarray}
where
\begin{eqnarray}
a(\vec{u},\vec{v}) &=& \int_{\Omega_\textrm{pore}} \mu\nabla\vec{u}: \nabla\vec{v} \hspace{3mm}d{\Omega_\textrm{pore}} \\
c(\vec{u},\vec{v}) = c(\vec{U};\vec{u},\vec{v}) &=& \int_{\Omega_\textrm{pore}}\rho(\vec{U}\cdot\nabla\vec{u})\cdot\vec{v}\hspace{3mm}d{\Omega_\textrm{pore}} \\
b(\vec{v},q) &=& -\int_{\Omega_\textrm{pore}} q \nabla\cdot \vec{v}\hspace{3mm}d{\Omega_\textrm{pore}}.
\end{eqnarray}

The solution to problem (\ref{eq:main}) can then be approximated as linear combination of multiscale basis functions $\vec{\Phi}_{E,j}={\Phi}_{E}\vec{e}_j$, $j = 1,2$ with $\{\vec{e}_{1},\vec{e}_{2}\}$ being the canonical basis of $\mathcal{R}^{2}$, such that:
\begin{equation}
\label{eq:expansion}
\vec{u}_H(x,y)=\sum_{E\in\mathcal{E}(H),\,j=1,2}{u}_{E,j}\vec{\Phi}_{E,j}(x,y).
\end{equation}
Consistent with the theory of MsFE method, the basis functions $\vec{\Phi}_{E,j}$ are themselves computed in the fine mesh constructed in each coarse elements. 
\subsection{Construction of a Crouzeix-Raviart basis}
For each edge $E\in \mathcal{E}_{H}$ we construct $\vec{\Phi}_{E,i}\in
V_{H}$, such that $\int_{E}\vec{\Phi}_{E,i}=\vec{e}_{i}$,
and $\int_{E^{\prime }}\vec{\Phi}_{E,i}=0$ for all $E^{\prime }\in \mathcal{E%
}_{H}$, $E^{\prime }\not=E$. 
These functions form a basis of $V_{H}$, i.e%
\begin{equation}\label{VHspan}
V_{H}=\textrm{span}\{\vec{\Phi}_{E,i},~E\in \mathcal{E}_{H},~i=1,2\}.
\end{equation}%
We also define $\textrm{supp}(\vec{\Phi}_{E,i})\subset \omega _{E}$, the ensemble of two quadrangles in $\mathcal{T}_{H}$ which share an edge $E$. Hence we solve for each coarse element $T_k$, a total of eight basis functions $\vec{\Phi}
_{E,i}$ (and consequentially $\pi
_{E,i}$) which are the solutions to:
\begin{eqnarray}
\label{eq:crstrong}
\mu\nabla^2 \vec{\Phi}_{E,i}+\rho(\vec{U}\cdot\nabla\vec{\Phi}_{E,i})-\nabla \pi _{E,i} &=&0,\textrm{ in }\Omega_\textrm{pore}\cap T_{k}, \\
\nabla\cdot\vec{\Phi}_{E,i} &=&\textrm{constant, in }\Omega_\textrm{pore}\cap T_{k}, \nonumber\\
\vec{\Phi}_{E,i} &=&0,\textrm{ in }\Omega_\textrm{grain}\cap T_{k}, \nonumber\\
\nabla \vec{\Phi}_{E,i}n-\pi _{E,i}n &=&\textrm{constant, on }F\cap \Omega_\textrm{pore}, \forall F\in \mathcal{E}(T_{k}), \nonumber\\
\int_{F}\vec{\Phi}_{E,i} &=&\left\{ 
\begin{array}{c}
\vec{e}_{i},~F=E \nonumber\\ 
0,~F\not=E%
\end{array}%
\right. ,\forall F\in \mathcal{E}(T_{k}), \nonumber\\
\int_{\Omega_\textrm{pore}\cap T_{k}}\pi _{E,i} &=&0.\nonumber
\end{eqnarray}

To solve equation (\ref{eq:crstrong}), we use Q1-Q1 finite element spaces in which both velocity and pressure degrees of freedom are defined on the same set of grid points. This arrangement is chosen due to the ease of programming and the computational efficiency. A stabilisation method is however necessary when this approach is considered. In a homogeneous Cartesian coordinate with fine element width of $h$, a stable solution can be achieved by perturbing the condition $\nabla \cdot \vec{\Phi}_{E,i} = 0$ with a pressure Laplacian term (see \cite{brezzifortin}). In a weak form, equation (\ref{eq:crstrong}) reduces to finding $\vec{\Phi}_{E,i}\in H^{1}(T_{k}\cap\Omega_\textrm{pore})$, $\pi _{E,i}\in L^2_{0}(T_{k}\cap \Omega_\textrm{pore})$, and the Lagrange multipliers $%
\vec{\lambda}_{F}$, $\forall F\in \mathcal{E}(T_{k})$, where $\mathcal{E}(T_{k})$ is the set of all the edges of the quadrangle $T_k$, satisfying
\begin{eqnarray}
\int_{T_k} \mu \nabla \vec{\Phi}_{E,i} :\nabla \vec{v}_h 
+\int_{T_k} \rho (\vec{U}\cdot\vec{\Phi}_{E,i})\cdot\vec{v}_h
-\int_{T_k}\pi _{E,i}\nabla\cdot\vec{v}_h \\+\sum_{F\in \mathcal{E}(T_{k})}\vec{\lambda}_{F}\cdot \int_{F}\vec{\Phi}_{E,i} =0,\nonumber\\
-\int_\Omega q_h\nabla\cdot\vec{\Phi}_{E,i}-\theta h^2 \int_\Omega\nabla \pi_{E,i} \cdot\nabla q_h = 0,\nonumber\\
\sum_{F\in \mathcal{E}(T_{k})}\vec{\mu}_{F}\cdot \int_{F}\vec{\Phi}_{E,i} =\vec{\mu}_{E}\cdot \vec{e}_{i},
\quad\forall \vec{\mu}_{F}\in \textrm{R}^{2},~F\in\mathcal{E}(T_{k}).\nonumber
\end{eqnarray}
Here, $\vec{v}_h$ and $q_h$ occupy the same finite element spaces as $\vec{\Phi}_{E,i}$ and $\pi_{E,i}$ respectively. $\theta$ is the stabilisation parameter which we set as $0.01$ for all our simulations (see \cite{brezzipitkaranta}). 

\begin{figure}[hbt]
	\centering
	\includegraphics[width = 0.75\linewidth]{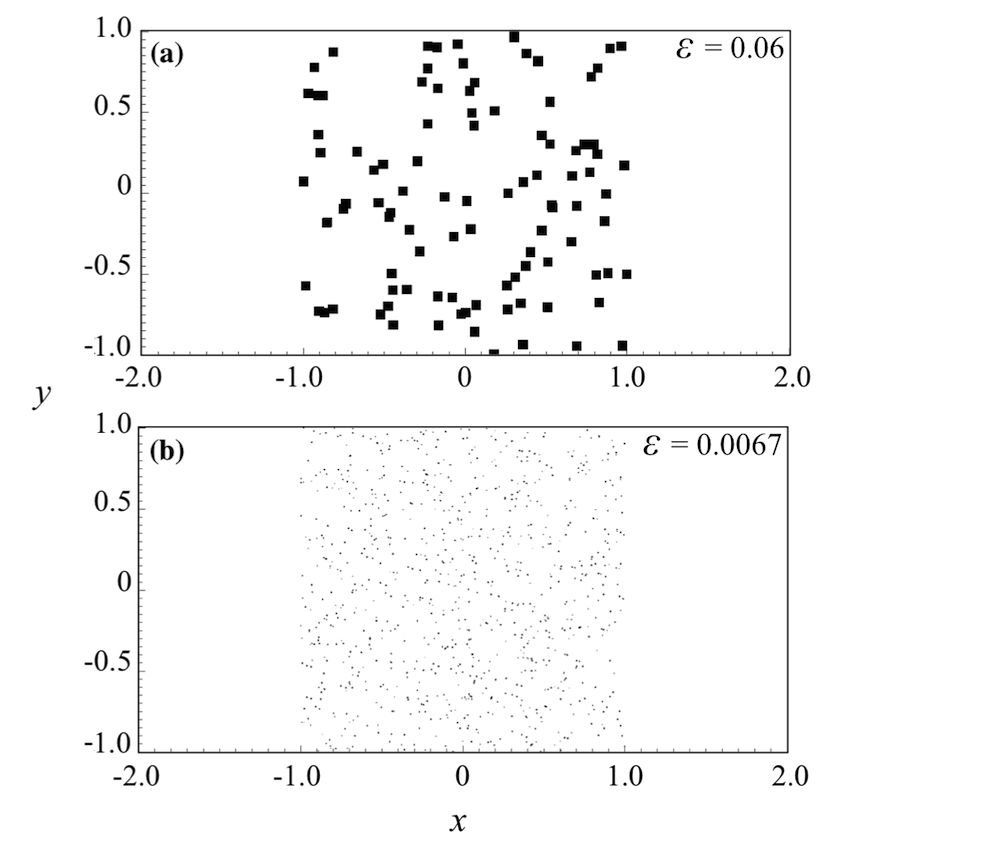} 
	\caption{Two grain patterns in a channel consisting of randomly laid (a) 100 rectangular grains with width $\varepsilon = 0.06$; and (b) 900 rectangular grains with $\varepsilon = 0.0067$.}
	\label{fig:grid}
\end{figure}
%%%%%%%%%%%%%%%%%%%%%%%%%%%%%%%%%%%%%%%%%%%%%%%%%%%%%%%%%%%%%%%%%%
%%%%%%%%%%%%%%%%
\begin{figure}[hbt]
	\centering
	\includegraphics[width = 0.9\linewidth]{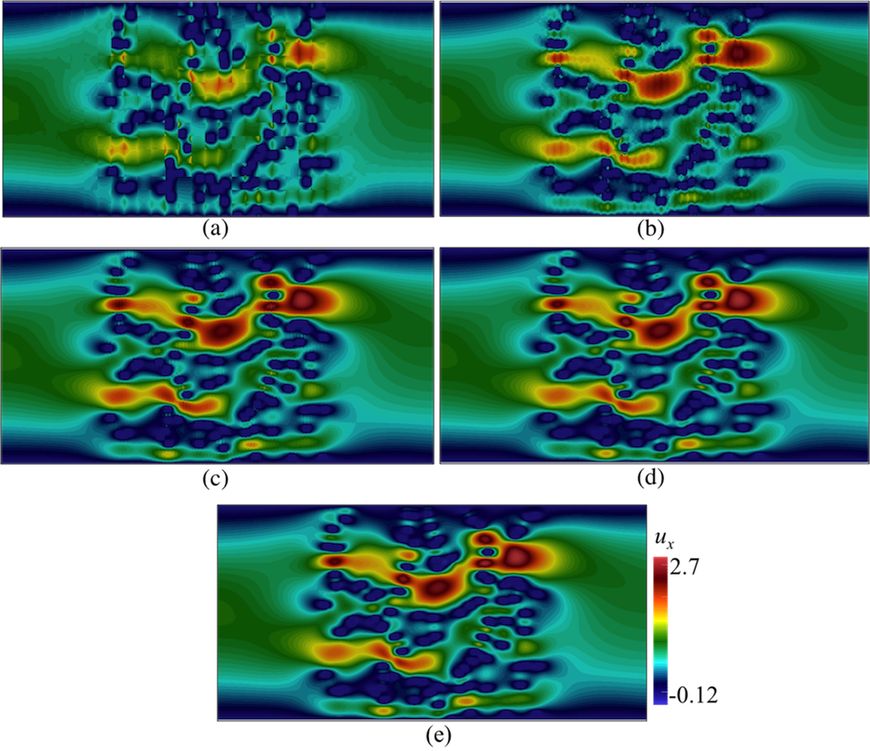} 
	\caption{Contours of $u_x$ in a domain depicted in figure \ref{fig:grid}(a), with $\vec{U}=(0.002,0)$ computed using Crouzeix-Raviart MsFEM on (a) $32\times16$; (b) $64\times 32$; (c) $128\times 64$; (d) $256\times128$ coarse elements; and (e) the reference solution.}
	\label{fig:channelGrainUx}
\end{figure}
%%%%%%%%%%%%%%%%%%%%%%%%%%%%%%%%%%%%%%%%%%%%%%%%%%%%%%%%%%%%%%%%%%%%%%%%%%%%%%%%%%
%%%%%%%%%%%%%%%%
\begin{figure}[hbt]
	\centering
	\includegraphics[width = 0.9\linewidth]{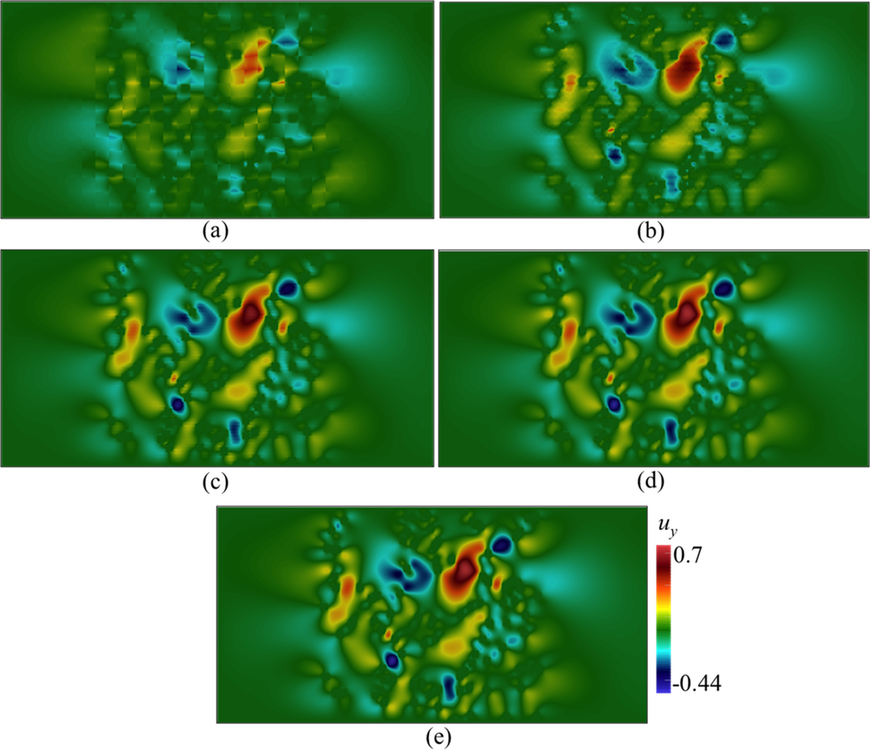} 
	\caption{Contours of $u_y$ in a domain depicted in figure \ref{fig:grid}(a), with $\vec{U}=(0.002,0)$ computed using Crouzeix-Raviart MsFEM on (a) $32\times16$; (b) $64\times 32$; (c) $128\times 64$; (d) $256\times128$ coarse elements; and (e) the reference solution.}
	\label{fig:channelGrainsUy}
\end{figure}
%%%%%%%%%%%%%%%%%%%%%%%%%%%%%%%%%%%%%%%%%%%%%%%%%%%%%%%%%%%%%%%%%%%%%%%%%%%%%%%%%%
\begin{figure}[hbt]
	\centering
	\includegraphics[width = 0.9\linewidth]{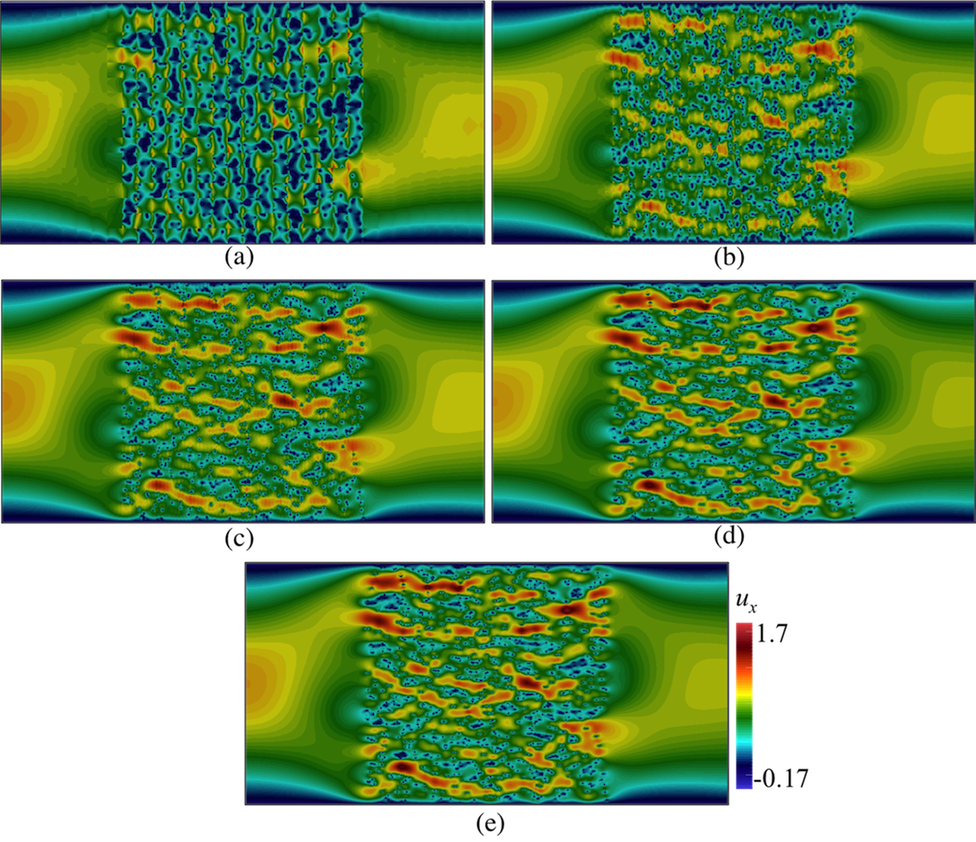} 
	\caption{Contours of $u_x$ in a domain depicted in figure \ref{fig:grid}(b), with $\vec{U}=(0.002,0)$ computed using Crouzeix-Raviart MsFEM on (a) $32\times16$; (b) $64\times 32$; (c) $128\times 64$; (d) $256\times128$ coarse elements; and (e) the reference solution.}
	\label{fig:channelDenseUx}
\end{figure} 
%%%%%%%%%%%%%%%%%%%%%%%%%%%%%%%%%%%%%%%%%%%%%%%%%%%%%%%%%%%%%%%%%%%%%%%%%%%%%%%%%%
%%%%%%%%%%%%%%%%%%%%%%%%%%%%%%%%%%%%%%%%%%%%%%%%%%%%%%%%%%%%%%%%%%%%%%%%%%%%%%%%%%
\begin{figure}[hbt]
	\centering
	\includegraphics[width = 0.9\linewidth]{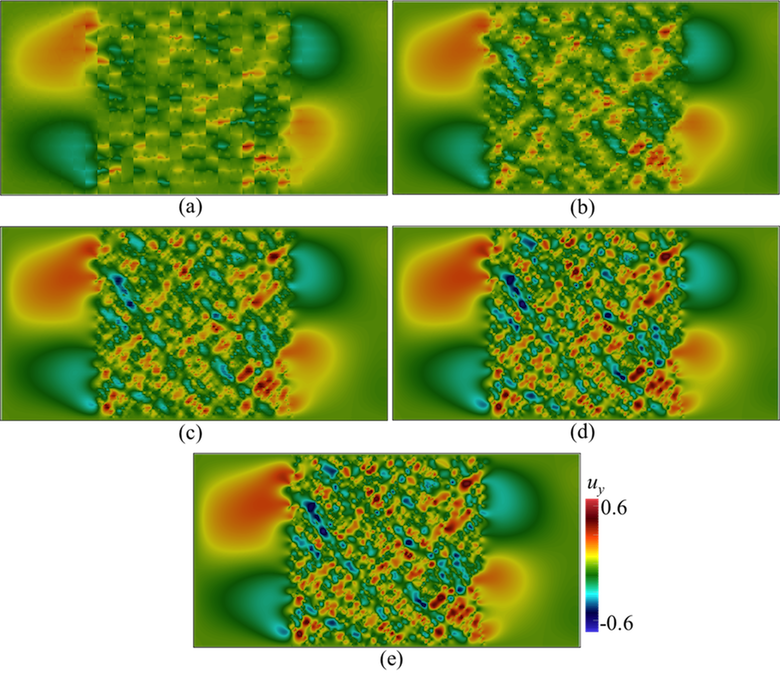} 
	\caption{Contours of $u_y$ in a domain depicted in figure \ref{fig:grid}(b), with $\vec{U}=(0.002,0)$ computed using Crouzeix-Raviart MsFEM on (a) $32\times16$; (b) $64\times 32$; (c) $128\times 64$; (d) $256\times128$ coarse elements; and (e) the reference solution.}
	\label{fig:channelDenseUy}
\end{figure} 
%%%%%%%%%%%%%%%%%%%%%%%%%%%%%%%%%%%%%%%%%%%%%%%%%%%%%%%%%%%%%%%%%%%%%%%%%%%%%%%%%%
\begin{figure}[hbt]
	\centering
	\includegraphics[width = 0.6\linewidth]{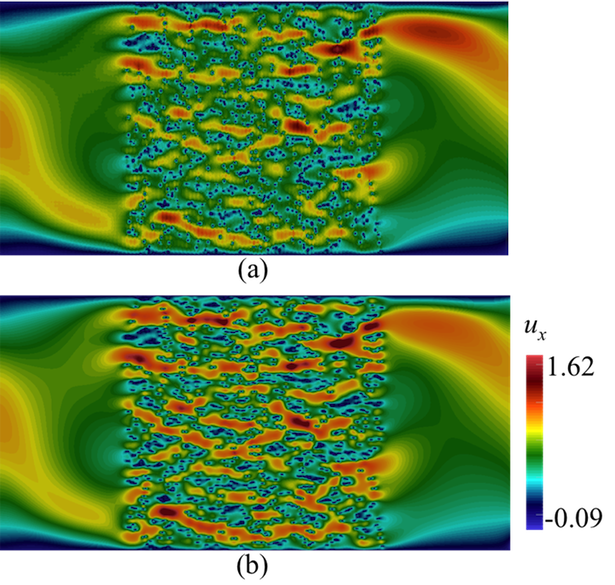} 
	\caption{Contours of $u_x$ in a domain depicted in figure \ref{fig:grid}(b), with $\vec{U}$ according to equation (\ref{eq:wind}), computed using Crouzeix-Raviart MsFEM on (a) $128\times 64$ coarse elements; compared with (b) the reference solution.}
	\label{fig:rotaWindux}
\end{figure} 
\begin{figure}[hbt]
	\centering
	\includegraphics[width = 0.6\linewidth]{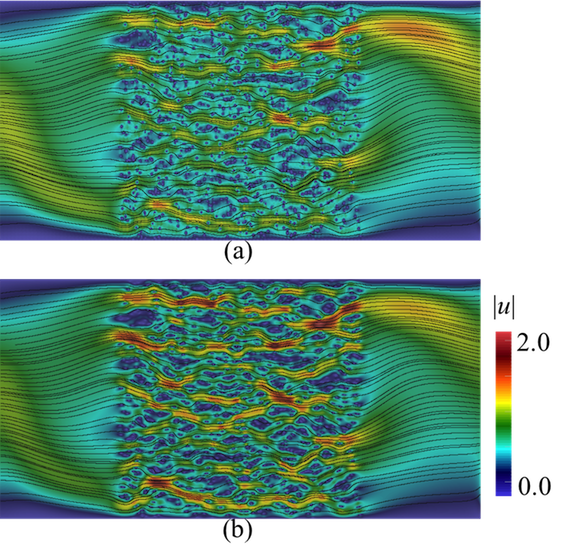} 
	\caption{Contours of magnitude of velocity $|u|$ along with their streamlines in a domain depicted in figure \ref{fig:grid}(b), with $\vec{U}$ according to equation (\ref{eq:wind}), computed using Crouzeix-Raviart MsFEM on (a) $128\times 64$ coarse elements; compared with (b) the reference solution.}
	\label{fig:streamRot}
\end{figure}
\begin{figure}[hbt]
	\centering
	\includegraphics[width = 0.9\linewidth]{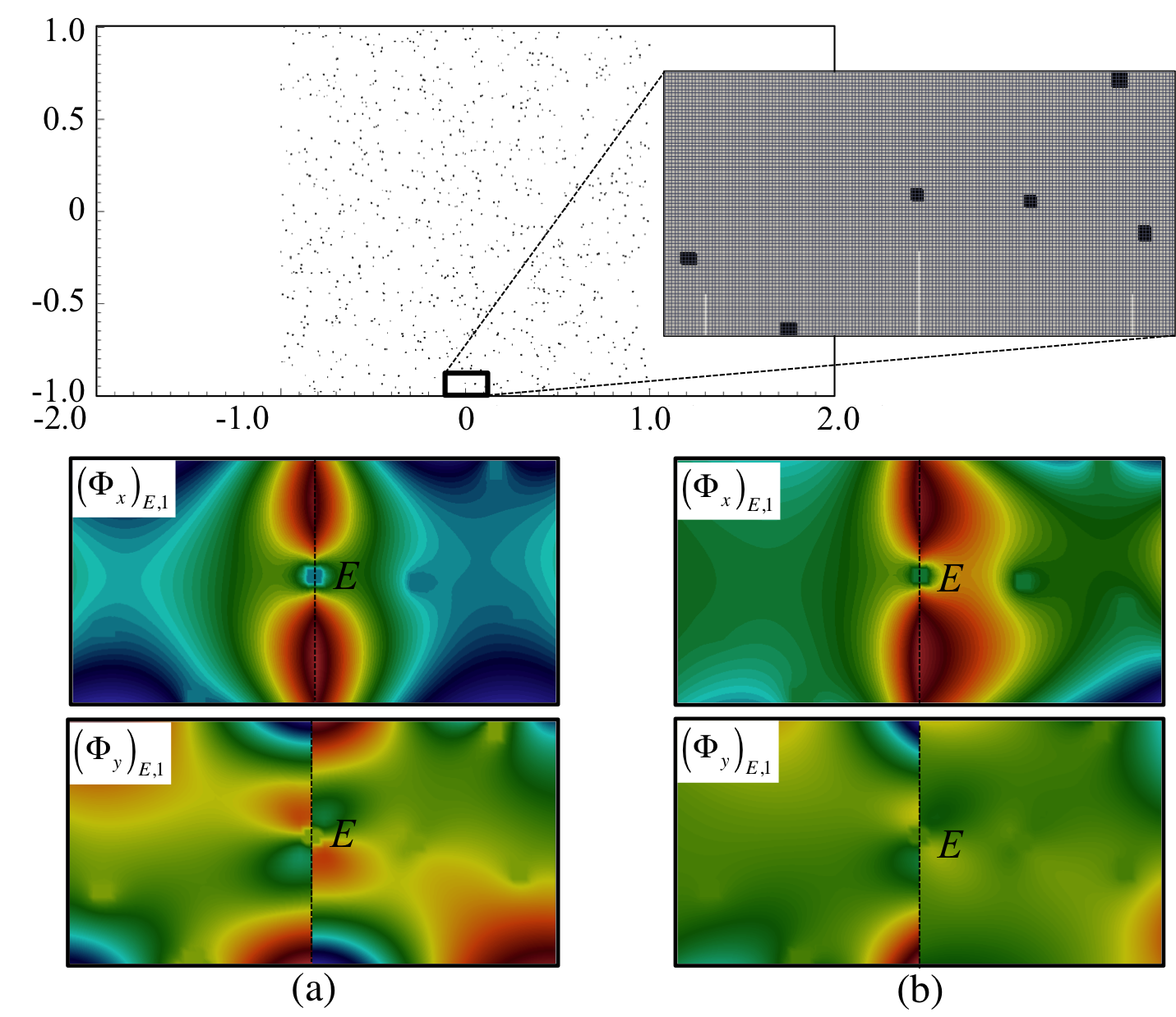} 
	\caption{Multiscale basis functions $\vec{\Phi}_{E,j},j=1,2$ in the presence of grains of which, one coincides with the edge $E$. The basis functions depicted are for the given velocity field (a) $\vec{U}=0$; and (b) according to equation (\ref{eq:wind}). In all cases, they successfully adapt to the worst-case scenario and maintain $\int_E \vec{\Phi}_{E,i}=\vec{e}_1$.}
	\label{fig:basis}
\end{figure}

\section{Penalization method}
\label{sec:penalization}
Often $\Omega_\textrm{pore}$ is a complicated pore structure in which solving equation (\ref{eq:main}) may require a complicated boundary-fitted, or even an unstructured mesh. In order to confine our computations in a homogeneous Cartesian mesh, we employ the penalization method \cite{Bruneau}. Henceforth, instead of solving equation (\ref{eq:main}) directly in $\Omega_\textrm{pore}$, we solve:
\begin{eqnarray}
\mu_\kappa \nabla^2 \vec{u}+\rho_\kappa(\vec{U}\cdot\nabla\vec{u}) + \sigma_\kappa\vec{u} - \nabla p &=& \vec{f}_\kappa \textrm{, in }  \Omega  \label{penal}\\
\nabla\cdot\vec{u}&=& 0 \textrm{, in }\Omega  \nonumber\\
\vec{u}&=&\vec{g}  \textrm{, on }  \partial\Omega \nonumber
\end{eqnarray}
in which
\begin{eqnarray}
\mu_\kappa;\rho_\kappa = \left\{ \begin{array}{cc}
\frac{1}{h} &\mbox{ in $\Omega_\textrm{grain}$} \\
\mu;\rho &\mbox{ in $\Omega_\textrm{pore}$}
\end{array}  \right.,
\sigma_\kappa = \left\{ \begin{array}{cc}
\frac{1}{h^3} &\mbox{ in $\Omega_\textrm{grain}$} \\
0 &\mbox{ in $\Omega_\textrm{pore}$}
\end{array}  \right.,
\vec{f}_\kappa = \left\{ \begin{array}{cc}
0 &\mbox{ in $\Omega_\textrm{grain}$} \\
\vec{f} &\mbox{ in $\Omega_\textrm{pore}$}
\end{array}  \right..
\end{eqnarray}
In our simulations the chosen fine-scale element width $h$ always satisfies $\varepsilon/h\geq5$. The penalization coefficient $\sigma_\kappa$ then forces the solution $\vec{u}$ to vanish inside the obstacles. Other variants of penalization methods are studied in \cite{Bruneau}.

\section{Reference solution}
\label{sec:reference}
We use a Q1-Q1 finite element method to compute the reference solutions, as we do for computing $\vec{\Phi}$, in the global fine mesh $\mathcal{T}_h$. In a weak form, the solution to equation (\ref{eq:main}) are $\vec{u}_h$ and $p_h$ such that
\begin{eqnarray}
\int_\Omega \mu_\kappa \nabla \vec{u}_h:\nabla\vec{v}_h +\int_\Omega\rho_\kappa(\vec{U}\cdot\nabla\vec{u}_h)\cdot\vec{v}_h +\int_\Omega \sigma_\kappa\vec{u}_h\cdot\vec{v}_h\\-\int_\Omega p_h\nabla\cdot\vec{v}_h = \int_\Omega \vec{v}_h \cdot \vec{f}_\kappa,\nonumber\\
-\int_\Omega q_h\nabla\cdot\vec{u}_h-\theta h^2(\nabla p_h,\nabla q_h) = 0,\nonumber\label{incomp}
\end{eqnarray}
where $\vec{v}_h$, and $p_h$ occupy Q1-Q1 finite element spaces. We use the stabilization parameter $\theta=0.01$ throughout this paper \cite{Elman}. Note that other stable or stabilized elements can be used to provide the reference solution.

\section{Boundary condition}
\label{sec:bc}
Note that the boundary condition $\vec{u}=\vec{g}$ on $\partial\Omega$ is not included in equation (\ref{eq:funcSpace}). This is possible since we approximate the boundary condition in equation \ref{eq:crstrong} only in a weak sense, i.e
\begin{equation}
\label{specialbc}
\int_{E} \vec{u}_H = \int_{E} \vec{g}, \quad\forall E \in \mathcal{E}_H \hspace{1mm}\textrm{on}\hspace{1mm}\partial\Omega.
\end{equation}
Together with equation (\ref{eq:expansion}), we apply on the boundary $\partial\Omega$:
\begin{equation}
u_{E,i} = \int_{E} {\vec{g}\cdot \vec{e}_i}.
\end{equation} 
This approach has been applied successfully in \cite{MuljadiSIAM} for the Stokes equation, and is a modification with respect to previous work \cite{clbetal,lozinskibubbleetal} where the boundary condition were strongly incorporated in the definition of $V_H$.  Our approach therefore gives more flexibility when implementing non zero $\vec{g}$, i.e every basis functions $\vec{\Phi}_{E,i}$ including those on boundary $\partial\Omega$ can be computed in the same fashion---according to equation (\ref{eq:crstrong}).
%%%%%%%%%%%%%%%%%%%%%%%%%%%%%%%%%%%%%%%%%%%%%%%%%%%%%%%%%%%%%%%%%%%%%%%%%%%%%%%%%%
\begin{table}[Hbt!]
	\begin{center}
		\begin{tabular}{{l}{c}{c}{c}}
			\hline
			$N_H$       & $L^1$ relative & $L^2$ relative & $H^1$ relative \\ 
			\hline
			$32\times 16$     & 0.13 & 0.142 & 0.222  \\
			$64\times 32$     & 0.08 & 0.088  & 0.172  \\
			$128\times 64$    & 0.052 & 0.060 & 0.089  \\
			$256\times 128$   & 0.012 & 0.013 & 0.07  \\
			\hline
		\end{tabular}
	\end{center}
	\caption{Convergence study of Poiseuille flow.}
	\label{tab:poiseuille}
\end{table}
%%%%%%%%%%%%%%%%%%%%%%%%%%%%%%%%%%%%%%%%%%%%%%%%%%%%%%%%%%%%%%%%%%%%%%%%%%%%%%%%%%
\begin{table}[Hbt!]
	\begin{center}
		\begin{tabular}{{l}{c}{c}{c}{c}}
			\hline
			$N_H$      &   $(H/\varepsilon)$  & $L^1$ relative & $L^2$ relative & $H^1$ relative \\ 
			\hline
			$32\times 16$   & 2.08  & 0.144 & 0.170 & 0.433  \\
			$64\times 32$  & 1.04  & 0.097 & 0.121 & 0.347  \\
			$128\times 64$  & 0.52  & 0.054 & 0.071 & 0.282  \\
			$256\times 128$ & 0.26  & 0.011 & 0.024 & 0.173  \\
			\hline
		\end{tabular}
	\end{center}
	\caption{Convergence study of channel flow pass pattern (a) with $\vec{U} = (0.002,0)$.}
	\label{tab:configA}
\end{table}
%%%%%%%%%%%%%%%%%%%%%%%%%%%%%%%%%%%%%%%%%%%%%%%%%%%%%%%%%%%%%%%%%%%%%%%%%%%%%%%%%%
\begin{table}[Hbt!]
	\begin{center}
		\begin{tabular}{{l}{c}{c}{c}{c}}
			\hline
			$N_H$      &   $(H/\varepsilon)$  & $L^1$ relative & $L^2$ relative & $H^1$ relative \\ 
			\hline
			$32\times 16$   & 18.74  & 0.304 & 0.331 & 0.71  \\
			$64\times 32$  & 9.37  & 0.157 & 0.155 & 0.542  \\
			$128\times 64$  & 4.68  & 0.082 & 0.097 & 0.482  \\
			$256\times 128$ & 2.34  & 0.037 & 0.049 & 0.239  \\
			\hline
		\end{tabular}
	\end{center}
	\caption{Convergence study of channel flow pass pattern (b) with $\vec{U} = (0.002,0)$.}
	\label{tab:configB}
\end{table}
%%%%%%%%%%%%%%%%%%%%%%%%%%%%%%%%%%%%%%%%%%%%%%%%%%%%%%%%%%%%%%%%%%%%%%%%%%%%%%%%%%
\begin{table}[Hbt!]
	\begin{center}
		\begin{tabular}{{l}{c}{c}{c}{c}}
			\hline
			$N_H$      &   $(H/\varepsilon)$  & $L^1$ relative & $L^2$ relative & $H^1$ relative \\ 
			\hline
			$32\times 16$   & 18.74  & 0.32 & 0.355 & 0.80  \\
			$64\times 32$  & 9.37  & 0.153 & 0.161 & 0.567  \\
			$128\times 64$  & 4.68  & 0.077 & 0.102 & 0.442  \\
			$256\times 128$ & 2.34  & 0.038 & 0.043 & 0.319  \\
			\hline
		\end{tabular}
	\end{center}
	\caption{Convergence study of channel flow pass pattern (b) with $\vec{U}$ according to equation (\ref{eq:wind}).}
	\label{tab:configBwind}
\end{table}
%%%%%%%%%%%%%%%%%%%%%%%%%%%%%%%%%%%%%%%%%%%%%%%%%%%%%%%%%%%%%%%%%%%%%%%%%%%%%%%%%%
\section{Numerical Results}
\label{sec:numerical}
We consider a channel domain $\Omega = [-2\leq x \leq 2, -1 \leq y \leq 1]$ containing a porous medium spanning from $x=-1$ to $x=1$. We then assign $\rho=1$, $\mu = 0.001$, and $\vec{f}=0$. At the inlet, the theoretical incompressible Poiseuille solution (parabolic velocity profile) is applied for all cases, i.e $\vec{u}=\left(1-y^2,0\right)$ on $x=-2$, whereas the Neumann boundary condition $\partial u/\partial n = 0$ is assumed at the outlet, $x=2$. No--slip boundary conditions at the top and bottom walls are applied. 

First we apply our method on simple Poiseuille flow without any porous bodies. Then two grain patterns are included, see figure \ref{fig:grid}. From here on we refer to them as pattern (a) depicted in figure \ref{fig:grid}(a), and pattern (b) depicted in figure \ref{fig:grid}(b). Pattern (a) consists of 100 randomly placed grains, each with width $\varepsilon=0.06$. Pattern (b) consists of 900 grains with $\varepsilon=0.0067$. 

We compare all our results to the reference solutions. When computing the reference solutions, we employ Q1-Q1 finite element method on a fine mesh consisting of $2560\times1280$ quadrangles. This ensures the ratio $\varepsilon/h\geq5$.

\subsection{Poiseuille flow}
We apply a vector field $\vec{U}=\left(0.002,0\right)$ corresponding to $\textrm{Re}\approx4$ where $\textrm{Re}=\frac{\rho\varepsilon |\vec{U}|}{\mu}$ (in the absence of grains, we assume that $\varepsilon$ is the channel diameter). In table \ref{tab:poiseuille}, the norms of error of the Crouzeix-Raviart MsFEM solutions relative to the reference solution on a number of coarse meshes are given, showing a convincingly converging trend.

\subsection{Pattern (a)}
Here we test our method in solving flow pass a porous body with a random pattern of grains. We consider pattern (a) where each grain is a rectangle with width $\varepsilon = 0.06$, $\textrm{Re}\approx0.12$. In figures \ref{fig:channelGrainUx} and \ref{fig:channelGrainsUy} the contours of velocity components $u_x$ and $u_y$ computed on a number coarse meshes are given. The results are compared to the reference solution. The contours computed using Crouzeix-Raviart MsFEM at $256\times128$ are identical to the reference solution; however, even at $32\times16$, the flow features already resemble those of the reference solution, and at $64\times32$ without any appreciable difference. Norms of error relative to the reference solution in $L^1, L^2$, and $H^1$ spaces are given in table \ref{tab:configA} showing a converging behaviour.

\subsection{Pattern (b)}
Here we simulate flow pass pattern (b) which contains finer ($\varepsilon = 0.0067$) and much more grains. We use the same kinds of coarse meshes, and vector field $\vec{U}$ as in the previous tests, which corresponds to $\textrm{Re}\approx0.013$. This is a more challenging test than the previous ones due to much larger $H/\varepsilon$ ratios---ranging from $18.74$ to $2.34$. This means each coarse elements in the vicinity of the grains has higher chances of being occupied by more than one grain, and therefore suffers more oscillations. 

In figures \ref{fig:channelDenseUx} and \ref{fig:channelDenseUy} the contours of $u_x$ and $u_y$ are given. As in the previous tests, the results are compared to the reference solution. The contours of ${u}_x$ and $u_y$ computed using Crouzeix-Raviart MsFEM at $256\times128$ are obviously identical to those of the reference solution. The results at $64\times32$ however already exhibit similar main flow features to the reference solution. This shows that the multiscale basis functions do well in capturing highly oscillatory fine-scale solutions. Similarly  monotonically decreasing relative error norms are shown in table \ref{tab:configB}.

\subsection{Heterogeneous velocity field $\vec{U}$}
To further test the feasibility of our method, we apply a heterogeneous velocity field
\begin{eqnarray}
\vec{U}=\left( \begin{array}{rl}
\begin{array}{cc}
2y(1-0.25 x^2)\\-x(1-y^2)
\end{array}
\end{array}\right)\textrm{ , in } \Omega_\textrm{pore}.
\label{eq:wind}
\end{eqnarray}
In figures \ref{fig:rotaWindux} the contours of $u_x$ through pattern (b) computed using Crouzeix-Raviart MsFEM on $128\times64$ coarse elements; and the reference solution are compared. The flow features are noticeably different than the previous results especially in regions away from the porous body. Indeed at the farthest from grains, the flow experiences  $\textrm{Re}\approx80$ where the inertial effect manifests the most. In table \ref{tab:configBwind}, the convergence study is given for a range of coarse meshes. Again we can see that our method gives good qualitative and quantitative agreement with the reference solution. In figures \ref{fig:streamRot} the contours of the magnitude of velocity computed using Crouzeix-Raviart MsFEM on $128\times64$ coarse elements are displayed along with their streamlines and compared with the reference solution. We notice that the Crouzeix-Raviart MsFEM gives an excellent agreement in terms of flow pattern with the reference solution.

In figures \ref{fig:basis}, we plot the multiscale basis function $\vec{\Phi}$ associated to the highlighted patch in the computational domain with pattern (b). We select this patch to illustrate the behaviour of $\vec{\Phi}$ at a worst-case scenario: a grain coinciding with a coarse element edge $E$. Figure \ref{fig:basis}(a) shows the basis functions $\vec{\Phi}_{E,1}$ computed with $\vec{U}=0$, the Oseen problem therefore reduces to a Stokes problem. Figure \ref{fig:basis}(b) shows the basis functions $\vec{\Phi}_{E,1}$ computed with $\vec{U}$ according to equation (\ref{eq:wind}). We notice the difference in the behaviour of $\vec{\Phi}$ due to different $\vec{U}$. In any cases the basis functions succesfully adapt to the presence of a grain at the edge $E$ and maintain $\int_E \vec{\Phi}_{E,i}=\vec{e}_i$.

\section{Concluding remarks}
\label{sec:concluding}
The Crouzeix-Raviart MsFEM has been developed and tested for solving Oseen's approximation for incompressible flow around solid bodies. The method performs very well in the presence of non-periodic grain formations. The weakly enforced continuity across coarse element edges ensures accurate basis function solutions without any oversampling methods. The basis functions are shown to successfully capture the effects of homogeneous and inhomogeneous vector field $\vec{U}$. The penalization method is seamlessly incorporated into our method allowing an extensive utilisation of simple Cartesian mesh.

This method is developed as a stepping stone towards solving then more complicated Navier-Stokes equation. Although only two-dimensional cases are considered, the extension of this work on three-dimensions is straightforward. Similarly the method can be applied for inhomogeneous Oseen's problem with $\vec{f}\neq0$. The reconstruction of fine-scale pressure is not the focus of the current work although it is possible (see \cite{MuljadiSIAM}). The calculations of MsFEM basis functions within a coarse element are done independent of the neighbouring elements which makes it suitable for the application of parallel programming.

For practical applications, the method is a promising development towards simulations capable of handling a wide range of spatial scales, while accommodating non-linear effects.

\section{Acknowledgement}
I thank the Engineering and Physical Science Research Council for financial support through grant number EP/L012227/1. I also thank Prof. Martin Blunt, and Prof. Pierre Degond for their invaluable advises. The source codes for the simulations in this paper are available at \url{https://www.imperial.ac.uk/engineering/departments/earth-science/research/research-groups/perm/research/pore-scale-modelling/software/}

\bibliographystyle{model1-num-names} 
\bibliography{springer}

\begin{thebibliography}{22}
\providecommand{\natexlab}[1]{#1}
\providecommand{\url}[1]{{#1}}
\providecommand{\urlprefix}{URL }
\expandafter\ifx\csname urlstyle\endcsname\relax
  \providecommand{\doi}[1]{DOI~\discretionary{}{}{}#1}\else
  \providecommand{\doi}{DOI~\discretionary{}{}{}\begingroup
  \urlstyle{rm}\Url}\fi
\providecommand{\eprint}[2][]{\url{#2}}

\bibitem[{Aarnes et~al(2005)Aarnes, Kippe, and Lie}]{Aarnes2005257}
Aarnes JE, Kippe V, Lie KA (2005) Mixed multiscale finite elements and
  streamline methods for reservoir simulation of large geomodels. Advances in
  Water Resources 28(3):257 -- 271,
  \doi{http://dx.doi.org/10.1016/j.advwatres.2004.10.007},
  \urlprefix\url{http://www.sciencedirect.com/science/article/pii/S0309170804001885}

\bibitem[{Angot et~al(1999)Angot, Bruneau, and Fabrie}]{Bruneau}
Angot P, Bruneau CH, Fabrie P (1999) A penalization method to take into account
  obstacles in incompressible viscous flows. Numerische Mathematik
  81(4):497--520, \doi{10.1007/s002110050401},
  \urlprefix\url{http://dx.doi.org/10.1007/s002110050401}

\bibitem[{Blunt et~al(2013)Blunt, Bijeljic, Dong, Gharbi, Iglauer, Mostaghimi,
  Paluszny, and Pentland}]{Blunt2013image}
Blunt MJ, Bijeljic B, Dong H, Gharbi O, Iglauer S, Mostaghimi P, Paluszny A,
  Pentland C (2013) {Pore-scale imaging and modelling}. Advances in Water
  Resources 51:197--216, \doi{10.1016/j.advwatres.2012.03.003},
  \urlprefix\url{10.1016/j.advwatres.2012.03.003}

\bibitem[{Brezzi and Fortin(1991)}]{brezzifortin}
Brezzi F, Fortin M (1991) Mixed and Hybrid Finite Element Methods.
  Springer-Verlag New York, Inc., New York, NY, USA

\bibitem[{Brezzi and Pitk{\"a}ranta(1984)}]{brezzipitkaranta}
Brezzi F, Pitk{\"a}ranta J (1984) Efficient Solutions of Elliptic Systems:
  Proceedings of a GAMM-Seminar Kiel, January 27 to 29, 1984. Vieweg+Teubner
  Verlag, Wiesbaden, \doi{10.1007/978-3-663-14169-3_2},
  \urlprefix\url{http://dx.doi.org/10.1007/978-3-663-14169-3_2}

\bibitem[{Bris et~al(2013)Bris, Legoll, and Lozinski}]{clbetal}
Bris C, Legoll F, Lozinski A (2013) {MsFEM} {\`a} la crouzeix-raviart for
  highly oscillatory elliptic problems. Chinese Annals of Mathematics, Series B
  34(1):113--138, \doi{10.1007/s11401-012-0755-7},
  \urlprefix\url{http://dx.doi.org/10.1007/s11401-012-0755-7}

\bibitem[{Bris et~al(2014)Bris, Legoll, and Lozinski}]{lozinskibubbleetal}
Bris CL, Legoll F, Lozinski A (2014) An msfem type approach for perforated
  domains. Multiscale Modeling \& Simulation 12(3):1046--1077,
  \doi{10.1137/130927826}, \urlprefix\url{http://dx.doi.org/10.1137/130927826}

\bibitem[{Chu et~al(2008)Chu, Efendiev, Ginting, and Hou}]{chuetal}
Chu J, Efendiev Y, Ginting V, Hou T (2008) Flow based oversampling technique
  for multiscale finite element methods. Advances in Water Resources 31(4):599
  -- 608, \doi{http://dx.doi.org/10.1016/j.advwatres.2007.11.005},
  \urlprefix\url{http://www.sciencedirect.com/science/article/pii/S030917080700173X}

\bibitem[{Chung et~al(0)Chung, Efendiev, Li, and Vasilyeva}]{Guanglian}
Chung ET, Efendiev Y, Li G, Vasilyeva M (0) Generalized multiscale finite
  element methods for problems in perforated heterogeneous domains. Applicable
  Analysis 0(0):1--26, \doi{10.1080/00036811.2015.1040988},
  \urlprefix\url{http://dx.doi.org/10.1080/00036811.2015.1040988}

\bibitem[{Crouzeix and Raviart(1973)}]{CRRairo}
Crouzeix M, Raviart PA (1973) Conforming and nonconforming finite element
  methods for solving the stationary stokes equations i. ESAIM: Mathematical
  Modelling and Numerical Analysis - Mod{\'e}lisation Math{\'e}matique et
  Analyse Num{\'e}rique 7(R3):33--75,
  \urlprefix\url{http://eudml.org/doc/193250}

\bibitem[{Degond et~al(2015)Degond, Lozinski, Muljadi, and
  Narski}]{muljadicrmsfem}
Degond P, Lozinski A, Muljadi BP, Narski J (2015) Crouzeix-raviart {MsFEM} with
  bubble functions for diffusion and advection-diffusion in perforated media.
  Communications in Computational Physics 17:887--907,
  \doi{10.4208/cicp.2014.m299},
  \urlprefix\url{http://journals.cambridge.org/article_S1815240615000237}

\bibitem[{Efendiev et~al(2013)Efendiev, Galvis, and Hou}]{efendievetal}
Efendiev Y, Galvis J, Hou TY (2013) Generalized multiscale finite element
  methods (gmsfem). J Comput Phys 251:116--135,
  \doi{10.1016/j.jcp.2013.04.045},
  \urlprefix\url{http://dx.doi.org/10.1016/j.jcp.2013.04.045}

\bibitem[{Elman et~al(2005)Elman, Silvester, and Wathen}]{Elman}
Elman H, Silvester DJ, Wathen AJ (2005) Finite Elements and Fast Iterative
  Solvers : with Applications in Incompressible Fluid Dynamics. Oxford
  University Press

\bibitem[{Henning and Peterseim(2013)}]{henningpeterseim}
Henning P, Peterseim D (2013) Oversampling for the multiscale finite element
  method. Multiscale Modeling \& Simulation 11(4):1149--1175,
  \doi{10.1137/120900332}, \urlprefix\url{http://dx.doi.org/10.1137/120900332}

\bibitem[{Hou and Wu(1997)}]{thomhou}
Hou TY, Wu XH (1997) A multiscale finite element method for elliptic problems
  in composite materials and porous media. Journal of Computational Physics
  134(1):169 -- 189, \doi{http://dx.doi.org/10.1006/jcph.1997.5682},
  \urlprefix\url{http://www.sciencedirect.com/science/article/pii/S0021999197956825}

\bibitem[{Jenny et~al(2003)Jenny, Lee, and Tchelepi}]{Jenny200347}
Jenny P, Lee S, Tchelepi H (2003) Multi-scale finite-volume method for elliptic
  problems in subsurface flow simulation. Journal of Computational Physics
  187(1):47 -- 67, \doi{http://dx.doi.org/10.1016/S0021-9991(03)00075-5},
  \urlprefix\url{http://www.sciencedirect.com/science/article/pii/S0021999103000755}

\bibitem[{Lipnikov et~al(2011)Lipnikov, Moulton, and Svyatskiy}]{Lipnikov}
Lipnikov K, Moulton JD, Svyatskiy D (2011) Adaptive strategies in the
  multilevel multiscale mimetic method for two-phase flows in porous media.
  Multiscale Modeling \& Simulation 9(3):991--1016, \doi{10.1137/100787544},
  \urlprefix\url{http://dx.doi.org/10.1137/100787544}

\bibitem[{Mostaghimi et~al(2012)Mostaghimi, Blunt, and
  Bijeljic}]{Mostaghimi2012}
Mostaghimi P, Blunt MJ, Bijeljic B (2012) Computations of absolute permeability
  on micro-ct images. Mathematical Geosciences 45(1):103--125,
  \doi{10.1007/s11004-012-9431-4},
  \urlprefix\url{http://dx.doi.org/10.1007/s11004-012-9431-4}

\bibitem[{Muljadi et~al(2015{\natexlab{a}})Muljadi, Blunt, Raeini, and
  Bijeljic}]{Muljadi2015}
Muljadi BP, Blunt MJ, Raeini AQ, Bijeljic B (2015{\natexlab{a}}) The impact of
  porous media heterogeneity on non-darcy flow behaviour from pore-scale
  simulation. Advances in Water Resources pp~--,
  \doi{http://dx.doi.org/10.1016/j.advwatres.2015.05.019}

\bibitem[{Muljadi et~al(2015{\natexlab{b}})Muljadi, Narski, Lozinski, and
  Degond}]{MuljadiSIAM}
Muljadi BP, Narski J, Lozinski A, Degond P (2015{\natexlab{b}}) Nonconforming
  multiscale finite element method for stokes flows in heterogeneous media.
  part i: Methodologies and numerical experiments. Multiscale Modeling \&
  Simulation 13(4):1146--1172, \doi{10.1137/14096428X},
  \urlprefix\url{http://dx.doi.org/10.1137/14096428X}

\bibitem[{Scheibe et~al(2015)Scheibe, Murphy, Chen, Rice, Carroll, Palmer,
  Tartakovsky, Battiato, and Wood}]{scheibe}
Scheibe TD, Murphy EM, Chen X, Rice AK, Carroll KC, Palmer BJ, Tartakovsky AM,
  Battiato I, Wood BD (2015) An analysis platform for multiscale hydrogeologic
  modeling with emphasis on hybrid multiscale methods. Groundwater
  53(1):38--56, \doi{10.1111/gwat.12179},
  \urlprefix\url{http://dx.doi.org/10.1111/gwat.12179}

\bibitem[{Wesseling(1992)}]{wesseling1992introduction}
Wesseling P (1992) An introduction to multigrid methods. Pure and applied
  mathematics, John Wiley \& Sons Australia, Limited,
  \urlprefix\url{https://books.google.co.uk/books?id=MznvAAAAMAAJ}

\end{thebibliography}

\end{document}